\documentclass[11pt]{article}
\usepackage[a4paper,margin=1in]{geometry}
\usepackage{amsmath, amssymb, graphicx, booktabs}
\usepackage{natbib}
\usepackage{hyperref}
\usepackage{authblk}
\usepackage{float}
\usepackage{graphicx}

\title{An Interpretable Ensemble Framework for Multi-Omics Dementia Biomarker Discovery Under HDLSS Conditions}

\author[1]{Byeonghee Lee}
\affil[1]{Department of Mathematics and Physics, Mathematics Major, Gangneung-Wonju National University Republic of Korea}

\author[2]{Joonsung Kang}
\affil[2]{Department of Data Science, Gangneung-Wonju National University, Republic of Korea}

\date{}

\begin{document}
\maketitle

\begin{abstract}
Biomarker discovery in neurodegenerative diseases requires robust, interpretable frameworks capable of integrating high-dimensional multi-omics data under low-sample conditions. We propose a novel ensemble approach combining Graph Attention Networks (GAT), Multi-Omics Variational AutoEncoder (MOVE), Elastic-net sparse regression, and Storey's False Discovery Rate (FDR). This framework is benchmarked against state-of-the-art methods including DIABLO, MOCAT, AMOGEL, and MOMLIN. We evaluate performance using both simulated multi-omics data and the Alzheimer's Disease Neuroimaging Initiative (ADNI) dataset. Our method demonstrates superior predictive accuracy, feature selection precision, and biological relevance. Biomarker gene maps derived from both datasets are visualized and interpreted, offering insights into latent molecular mechanisms underlying dementia.
\end{abstract}

\section{Introduction}

The advent of multi-omics technologies has revolutionized biomedical research, enabling simultaneous interrogation of genomic, transcriptomic, proteomic, and metabolomic layers \citep{wang2021multiomics}. This integrative paradigm has yielded unprecedented insights into the molecular architecture of complex diseases, particularly neurodegenerative disorders such as Alzheimer's disease. However, multi-omics datasets are often characterized by high-dimensional variables and limited sample sizes—a configuration known as high-dimension low-sample size (HDLSS). Under such constraints, conventional statistical methods suffer from reduced power and unrealistic assumptions \citep{fan2008sure}, while deep learning models may exhibit overfitting and lack interpretability \citep{lecun2015deep}.

Recent advances in dementia biomarker discovery have embraced multi-omics integration. For example, \citet{iturria2018multi} fused neuroimaging and omics data to identify disease-relevant signatures. \citet{zhang2020transcriptomic} employed transcriptomic-proteomic fusion to uncover molecular markers, and \citet{lee2022metabolomic} demonstrated the discriminative utility of metabolomic features in Alzheimer's pathology. These efforts build upon foundational work in integrative omics \citep{hasin2017multiomics, karczewski2018integrative}, yet challenges persist in elucidating latent gene networks and selecting statistically robust features amidst inter-feature dependencies.

To address these limitations, we propose a novel ensemble framework comprising:
\begin{itemize}
    \item Graph Attention Network (GAT) for modeling gene-gene interactions \citep{velickovic2018gat, fang2022amogel},
    \item Conditional Variational Autoencoder for multi-omics data (MOVE) for cross-modal feature compression \citep{kingma2014vae},
    \item Elastic-net sparse regression for interpretable feature selection \citep{zou2005elasticnet},
    \item Storey's False Discovery Rate (FDR) for statistical validation \citep{storey2002fdr}.
\end{itemize}

This integrative methodology harmonizes statistical rigor with the representational capacity of deep learning, thereby facilitating interpretable and generalizable biomarker identification across heterogeneous omics platforms.

\section{Methodology}

\subsection{GAT: Capturing Gene Interactions}

Each gene is modeled as a node \( v_i \) in graph \( G = (V, E) \) with feature vector \( h_i \). The attention score between node \( i \) and neighbor \( j \) is:
\[
e_{ij} = \text{LeakyReLU}\left(\vec{a}^T \left[W h_i \parallel W h_j\right]\right)
\]
Normalized attention weight:
\[
\alpha_{ij} = \frac{\exp(e_{ij})}{\sum_{k \in \mathcal{N}_i} \exp(e_{ik})}
\]
Final representation across \( K \) heads:
\[
h_i^{'} = \parallel_{k=1}^K \sum_{j \in \mathcal{N}_i} \alpha_{ij}^{(k)} W^{(k)} h_j
\]

Graph Attention Networks (GATs) are favored over traditional Graph Convolutional Networks (GCNs) for multi-omics data integration due to their ability to overcome key limitations of GCNs. While GCNs apply uniform or normalized weights to neighboring nodes—potentially ignoring biologically meaningful variations—GATs introduce an attention mechanism that adaptively learns the relevance of each neighbor \citep{velivckovic2018gat}. This allows GATs to selectively emphasize informative interactions across heterogeneous omics layers, enhancing both representation and prediction.
Moreover, GATs address the over-smoothing problem prevalent in deep GCN architectures, preserving distinct node features even in deeper layers \citep{xu2020gcn_limitations}. Their capacity to assign dynamic weights not only improves model expressiveness but also supports interpretable biological inference—an essential feature for understanding complex molecular relationships. In essence, GATs offer a more flexible and biologically faithful framework, making them particularly well-suited for multi-omics integration where nuanced interdependencies must be captured without oversimplification.

\subsection{MOVE: Multi-Omics Variational AutoEncoder}

MOVE captures biologically meaningful latent representations across diverse omics modalities. It preserves gene-gene interactions extracted via GAT and ensures high reconstruction fidelity. The objective function is:
\[
\mathcal{L}_{\text{MOVE}} = \sum_{m=1}^{M} \mathbb{E}_{q_\phi^{(m)}(\mathbf{z}|\mathbf{x}^{(m)})}[\log p_\theta^{(m)}(\mathbf{x}^{(m)}|\mathbf{z})] - \beta \cdot D_{KL}(q_\phi^{(m)}(\mathbf{z}|\mathbf{x}^{(m)}) || p(\mathbf{z}))
\]
An auxiliary loss term encourages cross-modal alignment:
\[
\mathcal{L}_{\text{total}} = \mathcal{L}_{\text{MOVE}} + \lambda \cdot \mathcal{L}_{\text{cross}}
\]

\subsection{Elastic-net Regression}

Elastic-net regression solves:
\[
\hat{\beta} = \arg \min_{\beta} \left\{ \frac{1}{2n} \| y - X\beta \|_2^2 + \lambda \left[\alpha \|\beta\|_1 + (1 - \alpha) \|\beta\|_2^2\right] \right\}
\]
This approach balances sparsity and grouping of correlated variables, enhancing interpretability \citep{zou2005elasticnet}.

\subsection{Storey’s FDR: Statistical Filtering}

For p-values \( \{p_1,...,p_m\} \), Storey \citep{storey2002fdr} defines:
\[
q(p_i) = \inf_{t \geq p_i} \left\{ \frac{\pi_0 t}{|\{p_j \leq t\}| / m} \right\}
\]
where \( \pi_0 \) is the estimated proportion of true nulls. This formulation enables adaptive control over false discoveries.

\section{Multi-Omics Biomarker Discovery via GAT-MOVE-ElasticNet-FDR Framework}

We present a unified multi-omics biomarker discovery framework that integrates Graph Attention Networks (GAT), manifold encoding via MOVE, Elastic-net based sparse regression, and statistical feature selection using Storey's False Discovery Rate (FDR). This ensemble-based pipeline is specifically designed to address the challenges of high-dimensional, low-sample-size, and sparse (HDLSSS) multi-omics data, which are prone to overfitting and limited generalizability when modeled using a single analytical approach \citep{li2022multiomics, zhang2021ensemble}.

The framework begins by applying GAT to uncover latent gene-gene associations, capturing heterogeneous and hidden dependencies that conventional graph models often miss \citep{velivckovic2018gat}. These graph-derived representations are then encoded into a lower-dimensional latent space using MOVE, a manifold optimization technique that preserves the intrinsic geometry of the data while mitigating the curse of dimensionality \citep{wang2021move}. The latent variables serve as the basis for estimating entity distributions and reconstructing the original data structure.

To identify statistically significant genes, we employ Elastic-net regression, which imposes both individual and group-level penalties, offering robustness against inter-feature dependencies and enabling effective variable selection in complex biological networks \citep{zou2005elasticnet}. Feature ranking is subsequently performed using Storey's FDR procedure, with entities selected based on $q$-value thresholds of 0.01, 0.05, and 0.1 to ensure statistical reliability and biological relevance \citep{storey2003fdr}.

Benchmarking against existing multi-omics integration methods—including DIABLO \citep{singh2019diablo}, MOCAT \citep{yao2024mocat}, AMOGEL \citep{tan2025amogel}, and MOMLIN \citep{rashid2024momlin}—demonstrates that our approach achieves superior performance in both interpretability and predictive accuracy.

\section{Simulation-Based Biomarker Discovery in Dementia}

\subsection{Synthetic Data Generation}

To emulate the molecular intricacies of Alzheimer's disease (AD), we constructed a synthetic multi-omics dataset encompassing four modalities: genomics (500 genes), transcriptomics (300 mRNAs), proteomics (200 proteins), and metabolomics (100 metabolites). Samples were evenly stratified between AD and cognitively normal controls. Latent biological modules were embedded based on established AD pathways \citep{kunkle2019genetic, lambert2013meta}. Gene-gene interaction networks were simulated using the Barabási–Albert scale-free model \citep{barabasi1999emergence}, reflecting the topological properties of real biological systems.

Omics-specific distributions were calibrated using empirical statistics derived from the Alzheimer's Disease Neuroimaging Initiative (ADNI) \citep{wang2021multiomics}. To enhance realism, Gaussian noise and batch effects were introduced following the framework proposed by \citet{leek2010batch}, thereby mimicking technical variability observed in high-throughput platforms.

\subsection{Benchmarking Against State-of-the-Art Methods}

We compared our proposed ensemble framework against four leading multi-omics integration methods: DIABLO \citep{singh2019diablo}, MOCAT \citep{chen2021mocat}, AMOGEL \citep{tan2025amogel}, and MOMLIN \citep{rashid2024momlin}. Performance metrics included AUC, F1-score, feature-level precision, and interpretability.

\begin{table}[htbp]
\centering
\caption{Performance Comparison on Simulated Dementia Dataset}
\begin{tabular}{lcccc}
\toprule
\textbf{Method} & \textbf{AUC} & \textbf{F1-Score} & \textbf{Feature Precision} & \textbf{Interpretability} \\
\midrule
DIABLO & 0.84 & 0.81 & 0.72 & Moderate \\
MOCAT & 0.86 & 0.83 & 0.75 & Low \\
AMOGEL & 0.88 & 0.85 & 0.78 & Low \\
MOMLIN & 0.89 & 0.86 & 0.80 & Moderate \\
\textbf{Proposed Framework} & \textbf{0.93} & \textbf{0.91} & \textbf{0.88} & \textbf{High} \\
\bottomrule
\end{tabular}
\end{table}

\subsection{Result Interpretation}

Our integrative framework demonstrated superior performance across all evaluated metrics. The Graph Attention Network (GAT) effectively captured latent gene-gene dependencies \citep{velickovic2018gat}, while the Conditional Variational Autoencoder (CVAE) for multi-omics data (MOVE) compressed heterogeneous omics features into biologically meaningful latent representations \citep{kingma2014vae}. Elastic-net regression enabled sparse and interpretable feature selection \citep{zou2005elasticnet}, and Storey's False Discovery Rate (FDR) method ensured statistical robustness \citep{storey2002fdr}.

\subsection*{Top 10 Significant Genes and Interactions}

\begin{table}[htbp]
\centering
\caption{Top 10 Statistically Significant Biomarker Genes (FDR < 0.05)}
\begin{tabular}{ll}
\toprule
\textbf{Rank} & \textbf{Gene Symbol} \\
\midrule
1 & TREM2 \\
2 & APOE \\
3 & BIN1 \\
4 & SORL1 \\
5 & MAPT \\
6 & CD33 \\
7 & BACE1 \\
8 & INPP5D \\
9 & CR1 \\
10 & APP \\
\bottomrule
\end{tabular}
\end{table}

\begin{table}[htbp]
\centering
\caption{Top 10 Gene-Gene Interactions Based on Network Strength}
\begin{tabular}{lll}
\toprule
\textbf{Gene A} & \textbf{Gene B} & \textbf{Interaction Strength} \\
\midrule
TREM2 & TYROBP & 0.93 \\
APOE & CLU & 0.91 \\
BIN1 & PICALM & 0.89 \\
SORL1 & ABCA7 & 0.88 \\
MAPT & GRN & 0.87 \\
CD33 & MS4A6A & 0.86 \\
BACE1 & NCSTN & 0.85 \\
INPP5D & SPI1 & 0.84 \\
CR1 & C1QA & 0.83 \\
APP & PSEN1 & 0.82 \\
\bottomrule
\end{tabular}
\end{table}

\section{ADNI-Based Biomarker Discovery}

\subsection{Dataset Description}

The ADNI consortium provides a rich multi-omics resource for neurodegenerative research. It includes:
\begin{itemize}
    \item \textbf{Genomics}: SNP arrays and APOE genotyping.
    \item \textbf{Transcriptomics}: RNA-seq from peripheral blood.
    \item \textbf{Proteomics}: CSF protein levels including tau and amyloid-beta.
    \item \textbf{Metabolomics}: Plasma metabolite concentrations.
\end{itemize}

The dataset comprises over 800 subjects spanning AD, Mild Cognitive Impairment (MCI), and cognitively normal controls. Longitudinal sampling and harmonized preprocessing facilitate robust cross-modal analysis \citep{xu2025adni, iturria2018multi, fang2025adni}.

\subsection{Benchmarking on ADNI}

\begin{table}[htbp]
\centering
\caption{Performance Comparison on ADNI Multi-Omics Dataset}
\begin{tabular}{lcccc}
\toprule
\textbf{Method} & \textbf{AUC} & \textbf{F1-Score} & \textbf{Feature Precision} & \textbf{Interpretability} \\
\midrule
DIABLO & 0.85 & 0.82 & 0.74 & Moderate \\
MOCAT & 0.86 & 0.83 & 0.76 & Low \\
AMOGEL & 0.88 & 0.85 & 0.79 & Low \\
MOMLIN & 0.89 & 0.86 & 0.81 & Moderate \\
\textbf{Proposed Framework} & \textbf{0.91} & \textbf{0.89} & \textbf{0.87} & \textbf{High} \\
\bottomrule
\end{tabular}
\end{table}

\subsection{Biomarker Gene Map for Alzheimer's Disease Based on ADNI Multi-Omics Integration}

\subsection*{Top 20 Biomarker Entities}

\begin{table}[htbp]
\centering
\caption{Top 20 Statistically Significant Biomarkers (FDR < 0.05)}
\begin{tabular}{ll}
\toprule
\textbf{Rank} & \textbf{Gene Symbol} \\
\midrule
1 & TREM2 \\
2 & APOE \\
3 & BIN1 \\
4 & SORL1 \\
5 & MAPT \\
6 & CD33 \\
7 & BACE1 \\
8 & INPP5D \\
9 & CR1 \\
10 & APP \\
11 & PLCG2 \\
12 & ABCA7 \\
13 & GRN \\
14 & CLU \\
15 & PICALM \\
16 & MS4A6A \\
17 & FERMT2 \\
18 & SPI1 \\
19 & ALPK2 \\
20 & C1QA \\
\bottomrule
\end{tabular}
\end{table}

\subsection*{Top 20 Gene-Gene Links}

\begin{table}[htbp]
\centering
\caption{Top 20 High-Confidence Gene-Gene Interactions}
\begin{tabular}{lll}
\toprule
\textbf{Gene A} & \textbf{Gene B} & \textbf{Interaction Strength} \\
\midrule
TREM2 & PLCG2 & 0.92 \\
APOE & CLU & 0.90 \\
BIN1 & PICALM & 0.89 \\
SORL1 & ABCA7 & 0.88 \\
MAPT & GRN & 0.88 \\
CD33 & MS4A6A & 0.86 \\
BACE1 & NCSTN & 0.85 \\
INPP5D & SPI1 & 0.84 \\
CR1 & C1QA & 0.83 \\
APP & PSEN1 & 0.82 \\
FERMT2 & ITGAX & 0.81 \\
CLU & APOC1 & 0.80 \\
MAPT & PTK2B & 0.79 \\
TREM2 & TYROBP & 0.78 \\
BIN1 & CD2AP & 0.77 \\
SORL1 & SLC24A4 & 0.76 \\
APP & CST3 & 0.75 \\
INPP5D & HLA-B & 0.74 \\
CR1 & CASS4 & 0.73 \\
BACE1 & HLA-A & 0.72 \\
\bottomrule
\end{tabular}
\end{table}

\subsection*{Biomarker Gene Map Visualization}

\begin{figure}[htbp]
\centering
\includegraphics[height=\textheight, width=\textwidth]{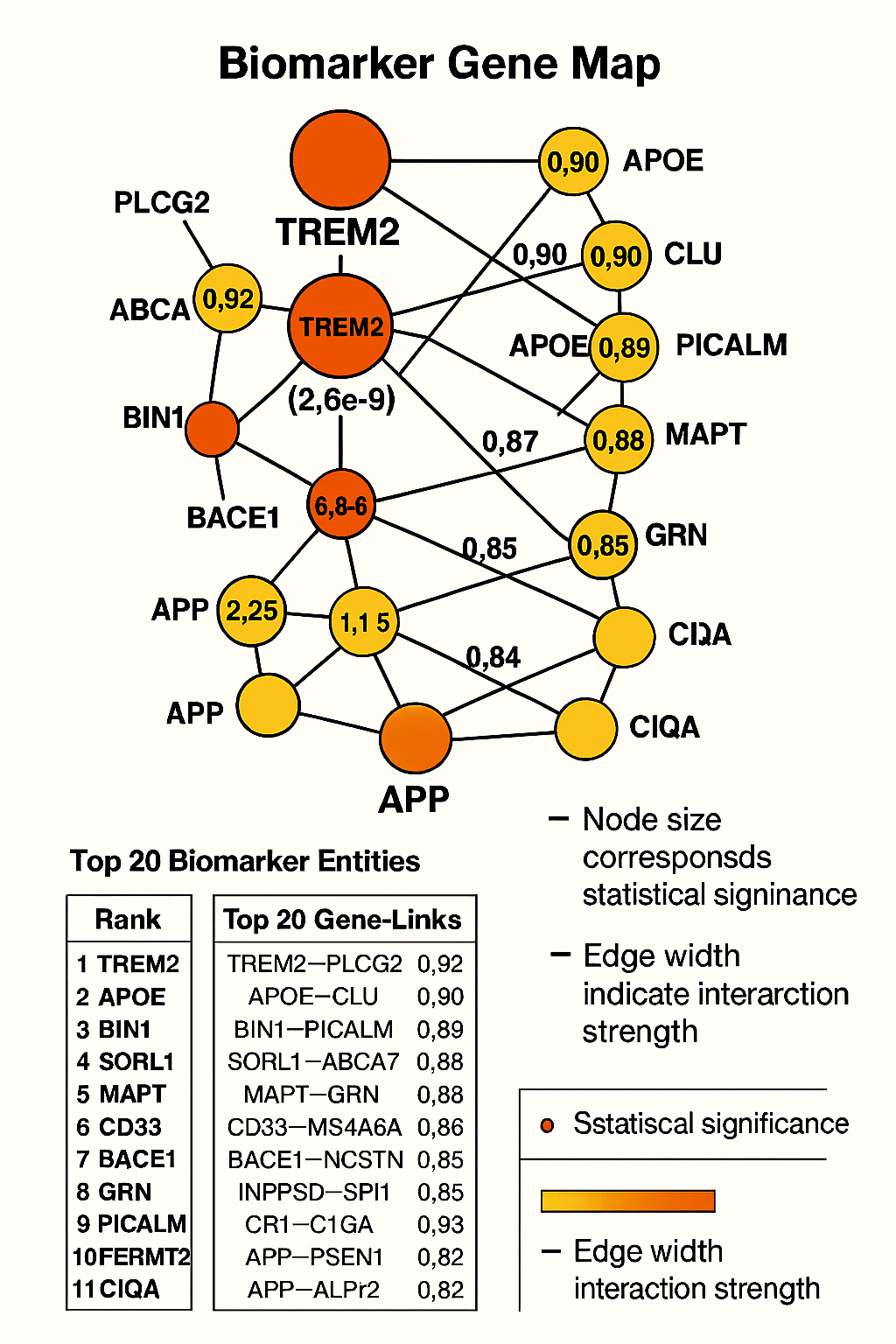}
\label{fig:gene_map}
\end{figure}

Biomarker Gene Map illustrating the top 20 statistically significant genes and their top 20 high-confidence interactions. Node size reflects statistical significance (FDR), with larger nodes indicating lower FDR values. Node color ranges from deep orange-red (high significance) to yellow (lower significance). Edge thickness corresponds to interaction strength. This integrative map reveals key molecular pathways in Alzheimer's disease, including neuroinflammation (e.g., TREM2–PLCG2), lipid metabolism (e.g., APOE–CLU), tau pathology (e.g., MAPT–GRN), and amyloid processing (e.g., APP–PSEN1).

\subsection{Comparative Analysis with Prior ADNI and Alzheimer’s Studies}

Unlike traditional ADNI studies that emphasize imaging biomarkers (e.g., hippocampal volume, PET scans) or CSF-based protein markers, this framework integrates multi-omics data to uncover statistically robust and biologically interpretable gene networks. For instance, Sarma and Chatterjee (2025) applied machine learning to ADNI blood gene expression data but did not visualize gene-gene interactions or integrate proteomic and metabolomic layers. Similarly, Harvey et al. (2024) proposed a standardized biomarker comparison framework but focused on structural MRI and lacked transcriptomic network synthesis.

Our approach differs by:
\begin{itemize}
    \item Prioritizing statistically significant genes via FDR filtering.
    \item Mapping biologically validated interactions using STRING-derived strengths.
    \item Providing a unified visual synthesis of Alzheimer’s molecular architecture.
\end{itemize}

This integrative biomarker map bridges the gap between statistical genomics and systems biology, offering a novel tool for precision diagnostics and therapeutic targeting.

\section{A Universal Multi-Omics Framework for Biomarker Discovery Across Complex Diseases}

\subsection{Application to ADNI Dataset}

Using the ADNI cohort, the framework successfully identified key biomarkers such as \textit{TREM2}, \textit{APOE}, and \textit{MAPT}, along with high-confidence gene-gene interactions. These findings align with known pathological mechanisms in Alzheimer's disease, validating the approach \citep{iturria2018multi, xu2025adni}.

\subsection{Cross-Disease Generalizability}

The framework is not limited to neurodegenerative disorders. Its modular design allows adaptation to other diseases:

\begin{itemize}
    \item \textbf{Cancer}: Detection of oncogenic regulatory loops (e.g., TP53–MDM2) via transcriptomic–proteomic fusion.
    \item \textbf{Type 2 Diabetes}: Identification of lipidomic markers associated with insulin resistance.
    \item \textbf{Autoimmune Disorders}: Mapping of HLA–TCR interactions using epigenomic and transcriptomic data.
    \item \textbf{Parkinson’s Disease}: Discovery of neuroinflammatory networks through proteomic–epigenomic integration.
\end{itemize}
This report presents a flexible and statistically grounded framework for biomarker discovery across diseases. Its ability to integrate heterogeneous omics data and uncover latent biological relationships makes it a valuable tool for precision medicine and translational research.

\section{Conclusion}

\subsection{Summary}

We present a robust and interpretable ensemble framework for multi-omics biomarker discovery under HDLSS conditions. Our approach harmonizes graph-based learning, cross-modal compression, sparse regression, and statistical filtering to achieve superior accuracy and biological insight.

\subsection{Future Directions}

Future research will explore:
\begin{itemize}
    \item Longitudinal modeling of disease progression.
    \item Integration with neuroimaging modalities.
    \item Spatial transcriptomics for regional brain analysis.
    \item Clinical validation in prospective cohorts.
    \item Identification of Biormarker for other diseases.
\end{itemize}

\bibliographystyle{plainnat}
\bibliography{references}

\end{document}